\documentclass[sigconf]{acmart}

\copyrightyear{2021}
\acmYear{2021}
\setcopyright{rightsretained}
\acmConference[CIKM '21]{Proceedings of the 30th ACM International Conference on Information and Knowledge Management}{November 1--5, 2021}{Virtual Event, QLD, Australia}
\acmBooktitle{Proceedings of the 30th ACM International Conference on Information and Knowledge Management (CIKM '21), November 1--5, 2021, Virtual Event, QLD, Australia}
\acmDOI{10.1145/3459637.3482124}
\acmISBN{978-1-4503-8446-9/21/11}

\usepackage{bm}
\usepackage{tabulary}
\usepackage{caption}
\usepackage{subcaption}

\definecolor{midnightgreen}{rgb}{0.0, 0.29, 0.33}

\begin{document}
\fancyhead[]{}

\title{Improving Query Representations for Dense Retrieval with Pseudo Relevance Feedback}

\author{HongChien Yu}
\affiliation{Carnegie Mellon University}
\email{hongqiay@cs.cmu.edu}

\author{Chenyan Xiong}
\affiliation{Microsoft Research}
\email{chenyan.xiong@microsoft.com}

\author{Jamie Callan}
\affiliation{Carnegie Mellon University}
\email{callan@cs.cmu.edu}
    
\begin{abstract}
Dense retrieval systems conduct first-stage retrieval using embedded representations and simple similarity metrics to match a query to documents. Its effectiveness depends on encoded embeddings to capture the semantics of queries and documents, a challenging task due to the shortness and ambiguity of search queries. This paper proposes ANCE-PRF, a new query encoder that uses pseudo relevance feedback (PRF) to improve query representations for dense retrieval. 
ANCE-PRF uses a BERT encoder that consumes the query and the top retrieved documents from a dense retrieval model, ANCE, and it learns to produce better query embeddings directly from relevance labels. It also keeps the document index unchanged to reduce overhead.  
ANCE-PRF significantly outperforms ANCE and other recent dense retrieval systems on several datasets. Analysis shows that the PRF encoder effectively captures the relevant and complementary information from PRF documents, while ignoring the noise with its learned attention mechanism.
\end{abstract}



\maketitle

\section{Introduction}

Dense retrieval systems first encode queries and documents into a dense embedding space and then perform information retrieval by finding a query's nearest neighbors in the embedding space~\cite{orqa, sentence-bert, dpr, ance}.
With the advancement of pre-trained language models~\cite{bert, lu2021less},  dedicated training strategies~\cite{dpr, ance}, and efficient nearest neighbor search~\cite{faiss, scann},
dense retrieval systems have shown effectiveness in a wide range of tasks, including web search~\cite{msmarco}, open domain question answering~\cite{naturalquestions}, and zero-shot IR~\cite{beir}.

Retrieval with dense, fully-learned representations has the potential to address some fundamental challenges in sparse retrieval. For example, vocabulary mismatch can be solved \textit{if} the embeddings accurately capture the information need behind a query and maps it to relevant documents. 
However, decades of IR research demonstrates that inferring a user's search intent from a concise and often ambiguous search query is challenging~\cite{croft2010search}.
Even with powerful pre-trained language models, it is unrealistic to expect an encoder to perfectly embed the underlying information need from a few query terms. 

A common technique to improve query understanding in sparse retrieval systems is 
\textit{pseudo relevance feedback} (PRF)~\cite{croft2010search,PRF,rm12}, which uses the top retrieved documents from an initial search as additional information to enrich the query representation.
Whether PRF information is used via query expansion~\cite{PRF,trec2004} or query term reweighting~\cite{bendersky2011parameterized}, its efficacy has been consistently observed across various search scenarios, rendering PRF a standard practice in many sparse retrieval systems.

This work leverages PRF information to improve query representations in dense retrieval.
Given the top retrieved documents from a dense retrieval model, e.g., ANCE~\cite{ance}, we build a PRF query encoder, ANCE-PRF, that uses a BERT encoder~\cite{bert} to consume the query and the PRF documents to refine the query representation. ANCE-PRF is trained end-to-end using relevance labels and learns to optimize the query embeddings using the rich information from PRF documents. It reuses the document index from ANCE to avoid duplicating index storage.

In experiments on MS MARCO and TREC Deep Learning (DL) Track passage ranking benchmarks, ANCE-PRF 
is consistently more accurate than ANCE and several recent dense retrieval systems that use more sophisticated models and training strategies~\cite{ltre,mebert}. 
We also observe large improvements on DL-HARD~\cite{dlhard} queries, a curated set to include complex search intents challenging for neural systems.
To the best of our knowledge, ANCE-PRF is among the best performing first-stage retrieval systems on the highly competitive MARCO passage ranking leaderboard.

Our studies confirm that the advantages of ANCE-PRF reside in its ability to leverage the useful information from the PRF documents while ignoring the noise from irrelevant PRF documents. 
The PRF encoder allocates substantially more attention to terms from the relevant PRF documents, compared to those from the irrelevant documents. 
A case study shows that the encoder focuses more on PRF terms that are complementary to the query terms in representing search intents.
These help ANCE-PRF learn better query embeddings that are closer to the relevant documents and improve the majority of testing queries.\footnote{Our code, checkpoints, and ranking results are open-sourced at \url{https://github.com/yuhongqian/ANCE-PRF}.}

\section{Related Work}

In \textit{dense retrieval} systems, queries and documents are encoded by a dual-encoder, often BERT-based, into a shared embedding space~\cite{dpr,ance,mebert,colbert}. 
Recent research in dense retrieval mainly focuses on improving the training strategies, especially the negative sampling part, including random sampling in batch~\cite{ance}, sampling from BM25 top negatives~\cite{orqa,clear}, sampling from an asynchronously~\cite{ance} updated hard negatives index, constructing hard negatives using document index from an existing dense retrieval model~\cite{ltre}, or reranking models~\cite{tas,tct}. 
Most dense retrieval systems encode a document using a constant number of embedding vectors~\cite{dpr,ance,mebert}, often one per document. There are also approaches using one vector per document token~\cite{colbert}, similar to the interaction-based neural IR approaches~\cite{drmm, knrm}. In this work, we focus on models that only use one vector per document, whose retrieval efficiency is necessary for real production systems~\cite{ance}.

In recent research, \textit{PRF} information has been leveraged by neural networks to combine feedback relevance scores~\cite{nprf}, modify query-document interaction using encoded feedback documents~\cite{dlcm,cobert}, or learn contextualized query-document interactions~\cite{pgt}. A parallel work~\cite{colbert-prf} expands multi-vector query representations with feedback embeddings extracted using a clustering technique.

\section{Method}

A typical dense retrieval system encodes query $q$ and document $d$ using a BERT-style encoder and then calculates the matching score $f(q, d)$ using simple similarity metrics: 
\begin{equation} \label{eq:score} 
    f(q, d) = \texttt{BERT}^q(\texttt{[CLS]} \circ q \circ \texttt{[SEP]}) \cdot \texttt{BERT}^d(\texttt{[CLS]} \circ d \circ \texttt{[SEP]}), 
\end{equation}
where BERT$^q$ and BERT$^d$ respectively output their final layer \texttt{[CLS]}  embeddings as the query and the document embeddings.
Eq. (\ref{eq:score}) is fine-tuned using standard ranking losses and with various negative sampling techniques~\cite{dpr, mebert}. 
The initial retrieval system this work uses, ANCE, conducts negative sampling from  an asynchronously updated document index~\cite{ance}.

ANCE-PRF leverages PRF documents retrieved by ANCE to enrich query representations. Given the top $k$ documents $d_1, ..., d_k$ from ANCE, ANCE-PRF trains a new PRF query encoder to output the query embedding $\bm{q^\text{prf}}$: 
\begin{equation} \label{eq:input}
    \bm{q^\text{prf}} = \texttt{BERT}^{prf}(\texttt{[CLS]} \circ q \circ \texttt{[SEP]} \circ d_1 \circ \texttt{[SEP]} \circ ... \circ d_k \circ \texttt{[SEP]}).
\end{equation}
It then conducts another retrieval with PRF embeddings: 
\begin{equation} \label{eq:prf_score}
    f^\text{prf}(q, d) = \bm{q^\text{prf}} \cdot \texttt{BERT}^d(\texttt{[CLS]} \circ d \circ \texttt{[SEP]}).
\end{equation}
The training uses the standard negative log-likelihood loss: 
\begin{equation} \label{eq:loss}
    \mathcal{L} = -\log \frac{\exp (\bm{q^\text{prf}} \cdot \bm{d^+})}{\exp (\bm{q^\text{prf}} \cdot \bm{d^+}) + \sum_{\bm{d^-} \in D^-} \exp (\bm{q^\text{prf}} \cdot \bm{d^-})},
\end{equation}
where $\bm{d^+}$ and $\bm{d^-}$ are embeddings of relevant and irrelevant documents. ANCE-PRF uses document embeddings from the initial dense retrieval model to avoid maintaining a separate document index for PRF. Therefore, only $\bm{q}^{prf}$ is newly learned. 

Eq. (\ref{eq:loss}) trains the query encoder to identify the relevant PRF information using its Transformer attention. Specifically, the attention from the \texttt{[CLS]} embedding in the last layer of Eq. (\ref{eq:input}) to the $j$th token $t_j$ of the input sequence $s$ is:
\begin{equation} \label{eq:cls_attn}
    cls\_attention(t_j) = \sum_i \frac{\exp(\bm{q}_{cls}^i \cdot \bm{k}_j^i)}{\sum_{l=1}^{|s|} \exp(\bm{q}_{cls}^i \cdot \bm{k}_l^i)}, 
\end{equation}
where $\bm{q}_{cls}^i$ and $\bm{k}_j^i$ are the ``query'' vector and $j$th input token's ``key'' vector of the $i$th attention head~\cite{transformer}.
Ideally, the PRF encoder should learn to yield
\begin{equation} \label{eq:goal}
    \sum_{j^+} cls\_attention(t_{j^+}) > \sum_{j^-} cls\_attention(t_{j^-}),
\end{equation}
where $j^+$ are indexes of the meaningful tokens from the PRF documents, and $j^-$ are those of the  irrelevant PRF tokens. 

ANCE-PRF can be easily integrated with any dense retrieval models.
With the document embeddings and index unchanged, the only computational overheads are one more query encoder forward pass (Eq. (\ref{eq:input})) and one more nearest neighbor search (Eq. (\ref{eq:prf_score})), a minor addition to the dense retrieval process~\cite{ance}.

\begin{table*}[t]
\footnotesize
\setlength{\tabcolsep}{2pt}
\caption{Ranking results. 
ANCE-PRF uses 3 feedback documents. All baseline results except BM25 and BM25+RM3 are reported by previous work. Statistically significant improvements over baselines are indicated by $\ast$ (BM25), $\dagger$ (BM25+RM3), $\ddagger$ (DE-BERT), $\mathsection$ (ME-BERT), and $\mathparagraph$ (ANCE) with $p \leq 0.05$ in t-test. Per query results of those \underline{underlined} are not available for significance tests. } \vspace{-3mm}
\begin{tabular}{l|l@{}r|l@{}r|l@{}r|l@{}r|l@{}r|l@{}r|c|l@{}r|l@{}r|c}
\hline 
\textbf{}         & \multicolumn{6}{c|}{\textbf{MARCO Dev}}             & \multicolumn{2}{c|}{\textbf{MARCO Eval}} & \multicolumn{5}{c|}{\textbf{TREC DL 2019}}           & \multicolumn{5}{c}{\textbf{TREC DL 2020}}           \\ \hline
\textbf{Method}   & \multicolumn{2}{c|}{\textbf{NDCG@10}} & \multicolumn{2}{c|}{\textbf{MRR@10}} & \multicolumn{2}{c|}{\textbf{R@1K}}  & \multicolumn{2}{c|}{\textbf{MRR@10}}     & \multicolumn{2}{c|}{\textbf{NDCG@10}} & \multicolumn{2}{c|}{\textbf{R@1K}}  & \textbf{HOLE@10} & \multicolumn{2}{c|}{\textbf{NDCG@10}} & \multicolumn{2}{c|}{\textbf{R@1K}}  & \textbf{HOLE@10} \\ \hline
\textbf{BM25}     & 0.238$^{\dagger}$ & -38.7\%       & 0.191$^{\dagger}$ & -42.1\%           & 0.858 & -10.5\%          & -   & -                & 0.506 & -21.9\%      & 0.750  & -0.1\%        & 0.000   & 0.480 & -25.7\%          & 0.786   & +1.3\%        & 0.006            \\
\textbf{BM25+RM3} & 0.219 & -43.6\%           & 0.171 & -48.2\%          & 0.872$^{\ast \mathsection}$ & -9.1\%           & -   & -                & 0.518  & -20.1\%           & \textbf{0.800}$^{\ast \mathparagraph}$ & +6.0\% & 0.000   & 0.482  & -25.4\%          & \textbf{0.822}$^{\ast  \mathparagraph}$ & +5.9\% & 0.002  \\
\textbf{DPR~\cite{dpr,ance}}      & -    & -            & \underline{0.311} & -5.8\%           & \underline{0.952}  & -0.1\%         & -  & -                 & \underline{0.600}  & -7.4\%          & -  & -             & -                & \underline{0.557}   & -13.8\%         & -              & -      & -          \\
\textbf{DE-BERT~\cite{mebert}}  & 0.358$^{\ast \dagger}$ & -7.7\%           & 0.302  & -8.5\%         & -   & -           & \underline{0.302}  & -4.7\%              & \underline{0.639}  & -1.4\%          & -  & -            & 0.165       & -     & -                & -     & -         & -                \\
\textbf{ME-BERT~\cite{mebert}}  & 0.394$^{\ast \dagger \ddagger}$ & +1.5\%           & 0.334$^{\ast \dagger \ddagger}$  & +1.2\%         & 0.855  & -10.8\%        & \underline{0.323}  & +1.9\%             & \underline{\textbf{0.687}}  & +6.0\%  & -  & -             & 0.109  & -          & -                & -              & -     & -           \\
\textbf{LTRe~\cite{ltre}}     & -  & -               & \underline{0.341}  & +3.3\%          & \underline{\textbf{0.962}} & +0.0\%         & -  & -                 & \underline{0.675}  & +4.2\%          & -  & -            & -                & -                & -              & -    & - & -            \\
\textbf{ANCE~\cite{ance}}     & 0.388$^{\ast \dagger \ddagger}$ & 0.0\%           & 0.330$^{\ast \dagger \ddagger}$ & 0.0\%         & 0.959$^{\ast \dagger \mathsection}$ & 0.0\%         & 0.317 & 0.0\%              & $0.648^{\ast \dagger}$  & 0.0\%          & 0.755  & 0.0\%         & 0.149            & $0.646^{\ast \dagger}$    & 0.0\%        & 0.776   & 0.0\%       & 0.135            \\ \hline
\textbf{ANCE-PRF} & \textbf{0.401}$^{\ast \dagger \ddagger \mathsection \mathparagraph}$ & +3.4\%     & \textbf{0.344}$^{\ast \dagger \ddagger \mathsection \mathparagraph}$ & +4.2\%  & 0.959$^{\ast \dagger \mathsection}$ & +0.0\% & \textbf{0.330} & +4.1\%      & $0.681^{\ast \dagger}$ & +5.1\%        & $0.791^{\mathparagraph}$   & +4.8\%       & 0.133            & $\textbf{0.695}^{\ast \dagger \mathparagraph}$  & +7.6\%  & 0.815  & +5.0\%         & 0.087            \\ \hline 
\end{tabular}

\label{tab:main_tab}
\end{table*}

\begin{table*}[t]
\footnotesize
\caption{Ranking accuracy with a varying number of PRF documents ($k$). 
$Avg\_Rel$ is the average relevance score of PRF documents at position $k$. Superscripts$^k$ mark statistically significant improvements over $k$. ANCE results are in the first row ($\ast$).} \vspace{-3mm}
\begin{tabular}{c|llll|llll|llll}
\hline 
           & \multicolumn{4}{c|}{\textbf{MARCO Dev} (Binary Label)}             & \multicolumn{4}{c|}{\textbf{TREC DL 2019} (0-3 Scale Label)}         & \multicolumn{4}{c}{\textbf{TREC DL 2020} (0-3 Scale Label)}          \\ \hline
\textbf{$\bm{k}$} & \textbf{NDCG@10} & \textbf{MRR@10} & \textbf{R@1K} & \textbf{Avg\_Rel}  & \textbf{NDCG@10}  & \textbf{R@1K} & \textbf{HOLE@10} & \textbf{Avg\_Rel} & \textbf{NDCG@10}  & \textbf{R@1K} & \textbf{HOLE@10} & \textbf{Avg\_Rel} \\ \hline
\textbf{$\ast$} &  0.388$^{0}$     &  0.330$^{0}$  &  0.959$^{0}$  & -  &  0.648     &   0.755    & 0.149  &  -   &    0.646    &   0.776     & 0.135 &   -   \\

\textbf{0} &  0.364     &  0.307  &  0.943  & - &  0.672     &   0.780    & 0.149   & -   &  0.668    &   0.791     & 0.115  & -   \\
\textbf{1} &  0.393$^{\ast0}$     &   0.334$^{0}$              & \textbf{0.963}$^{\ast0}$    &   0.210    &  0.680$^{\ast}$    &   0.795$^{\ast}$    &  0.142   & 2.023   &  0.689$^{\ast}$    &    0.814   & 0.093  &  2.093    \\
\textbf{2} & 0.401$^{\ast01}$       & 0.343$^{\ast01}$       &  0.962$^{03}$   & 0.112              & 0.678     & \textbf{0.797}$^{\ast}$  & 0.133  & 1.651 & \textbf{0.696}$^{\ast}$             & 0.816  & 0.085 & 1.870 \\
\textbf{3} & 0.401$^{\ast01}$            & 0.344$^{\ast01}$     & 0.959$^{0}$  & 0.067  & \textbf{0.681}     & 0.791$^{\ast}$ &  0.133   & 1.791  & 0.695$^{\ast}$  & 0.815   & 0.087   & 1.907   \\ 
\textbf{4} & \textbf{0.403}$^{\ast015}$   & \textbf{0.346}$^{\ast012}$   & 0.961$^{0}$    & 0.046 &  0.675         &  0.796$^{\ast}$  &  0.130  & 1.535 & \textbf{0.696}$^{\ast}$                  &  \textbf{0.821}   & 0.093  & 1.556 \\ 
\textbf{5}   & 0.400$^{\ast01}$    & 0.344$^{\ast01}$  & 0.960$^{01}$   & 0.039  &  \textbf{0.681}      & 0.796$^{\ast}$  &  0.128  & 1.465  &  0.688$^{\ast}$         &  0.816  & 0.096 & 1.370  \\ 
\hline 
\end{tabular}
\label{tab:ablation} 
\end{table*}

\begin{table}[t]
\footnotesize
\caption{Results on DL-HARD~\cite{dlhard}. We use the same symbols as in Table~\ref{tab:main_tab} for statistically significant improvements.}  \vspace{-3mm}
\begin{tabular}{l|lr|lr|c}
\hline 
\textbf{}         & \multicolumn{5}{c}{\textbf{DL-HARD}}                \\ \hline
\textbf{Method}   & \multicolumn{2}{c|}{\textbf{NDCG@10}} & \multicolumn{2}{c|}{\textbf{R@1K}}  & \textbf{HOLE@10} \\ \hline
\textbf{BM25}     & 0.304$^{\dagger}$ & -9.0\%           & 0.669   & -12.8\%       & 0.504   \\
\textbf{BM25+RM3} & 0.273   & -18.3\%         & 0.703   & -8.3\%       & 0.508            \\
\textbf{ANCE}     & 0.334   & 0.0\%           & \textbf{0.767}  & 0.0\% & 0.570            \\ \hline
\textbf{ANCE-PRF} & \textbf{0.365}$^{\dagger}$ & +9.3\%   & 0.761  & -0.1\%        & 0.544            \\ \hline 
\end{tabular}
\label{tab:dlhard}
\end{table}

\section{Experimental Setup}
\label{sec:baseline}
Next, we discuss the datasets, baselines, and implementation details. 

\textbf{Datasets.} We use \textit{MS MARCO} passage training data~\cite{msmarco} which includes 530K training queries. We first evaluate on its dev set with 7k queries and also obtain the testing results by submitting to its leaderboard.
MARCO's official metric is MRR@10. 

We also evaluate the MARCO trained model on two additional evaluation benchmarks, TREC DL~\cite{trecdl19,trecdl20} and DL-HARD~\cite{dlhard}.
{\textit{TREC DL}}~\cite{trecdl19,trecdl20} includes 43 labeled queries from 2019 and 54 from 2020 for the MARCO corpus.
The official metric is NDCG@10 and Recall@1K, the latter with label binarized at relevance point 2. Following Xiong et al.~\cite{ance}, 
we also report HOLE@10, the unjudged fraction of top 10 retrieved documents, to reflect the coverage of pooled labels on dense retrieval systems.
\textit{DL-HARD}~\cite{dlhard} contains 50 queries from TREC DL that were curated to challenge neural systems in a prior TREC DL track. Its official metric is NDCG@10. 

\textbf{Baselines} include  {\textit{BM25}}~\cite{bm25},  {\textit{RM3}}~\cite{trec2004,rm12}, a classical PRF framework in sparse retrieval. We also compare with several recent dense retrievers. 
{\textit{ME-BERT}}~\cite{mebert} was trained with hard-negative mining~\cite{deer}, and is the only one that uses multi-vector document encoding. 
{\textit{DE-BERT}}~\cite{mebert} is the single-vector version of ME-BERT. {\textit{DPR}}~\cite{dpr} is trained with in-batch negatives. {\textit{LTRe}}~\cite{ltre} generates hard negatives using document embeddings from an existing dense retrieval model. {\textit{ANCE}}~\cite{ance} uses 
hard negatives from asynchronously updated dense retrieval index using the latest model checkpoint.

\textbf{Implementation Details.} In training, we initialize query encoder from the ANCE FirstP model~\cite{ance}\footnote{\url{https://github.com/microsoft/ANCE}} and kept the document embeddings from ANCE (and thus also the ANCE negative index) uncharged.  
All hyperparameters used in ANCE training are inherited in ANCE-PRF.
All models are trained on two RTX 2080 Ti GPUs with per-GPU batch size 4 and gradient accumulation step 8 for 450K steps. We keep the model checkpoint with the best MRR@10 score on the MS MARCO dev set. 

In inference, we first obtain ANCE top $k$ documents using Faiss IndexFlatIP and Eq. (\ref{eq:score}), feed them into the ANCE-PRF query encoder (Eq. (\ref{eq:input})) for updated query embeddings, and run another Faiss search with Eq.~(\ref{eq:prf_score}) for final results.

\section{Experimental Results}
In this section, we discuss our experimental results and studies.

\subsection{Overall Results} 
Table~\ref{tab:main_tab} includes overall retrieval accuracy on MS MARCO and TREC DL datasets. 
ANCE-PRF  outperforms ANCE, its base retrieval system, on all datasets. 
On the challenging DL-HARD (Table~\ref{tab:dlhard}), ANCE-PRF improves NDCG@10 By 9.3\% over ANCE, indicating ANCE-PRF's advantage in queries challenging for neural systems. These results suggest that ANCE-PRF effectively leverages PRF information to produce better query embeddings. ANCE-PRF also helps retrieve relevant documents not recognized by ANCE, improving R@1K by about 5\% on both TREC DL sets.



ANCE-PRF rankings are significantly more precise than the sparse retrieval baselines with large margins across all datasets. RM3 achieves the best R@1K on both TREC DL sets, but its improvement is not as significant on DL-HARD.

ANCE-PRF also outperforms several strong dense retrieval baselines and produces the most accurate rankings on almost all datasets.
While Luan et al.~\cite{mebert} discuss the theoretical benefits of higher dimensional dense retrieval as in ME-BERT, our empirical results show that a well-informed query encoder can achieve comparable results, while avoiding the computational and spatial overhead caused by using multiple vectors per document.


\begin{figure*}[t]
    \centering
    \begin{subfigure}[b]{0.235\textwidth}
    \centering
    \includegraphics[width=\textwidth]{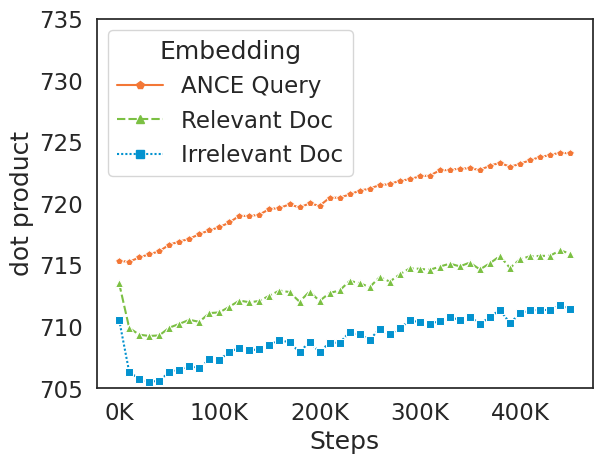}
    \caption{Dot product}
    \end{subfigure} 
    \begin{subfigure}[b]{0.235\textwidth}
    \centering
    \includegraphics[width=\textwidth]{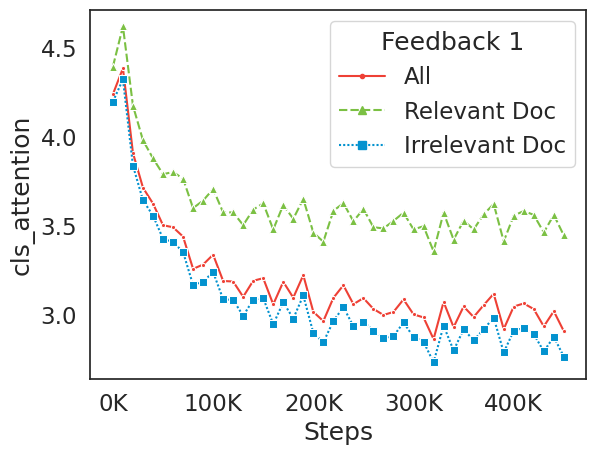}
    \caption{Attention on 1st PRF Doc}
    \end{subfigure} 
    \begin{subfigure}[b]{0.235\textwidth}
    \centering
    \includegraphics[width=\textwidth]{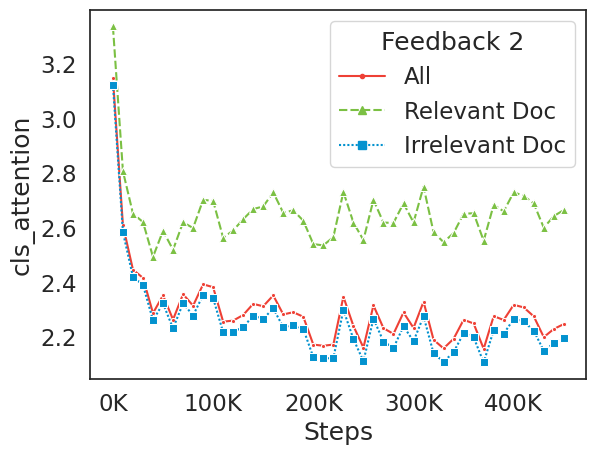}
    \caption{Attention on 2nd PRF Doc}
    \end{subfigure} 
        \begin{subfigure}[b]{0.235\textwidth}
    \centering
    \includegraphics[width=\textwidth]{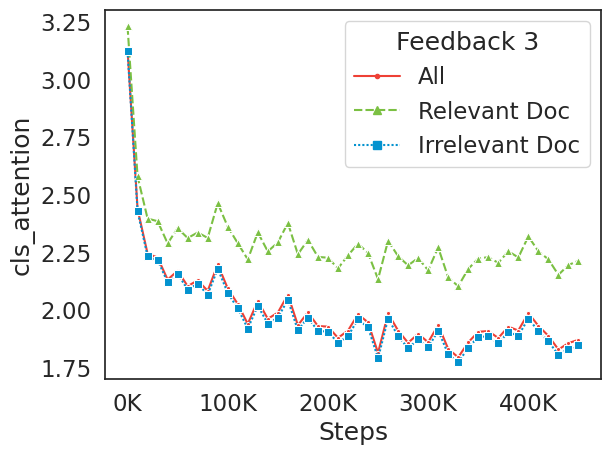}
    \caption{Attention on 3rd PRF Doc}
    \end{subfigure} 
    \caption{Fig. (a) shows the dot product between the ANCE-PRF query embedding and document embeddings at different training steps (x-axis). Fig. (b)-(d) are the cls$\_$attention (y-axes) on ``all'', ``relevant'', and ``irrelevant'' feedback documents ranked at positions 1-3 in the initial retrieval.  
    }
    \label{fig:dp_attn}
\end{figure*}

\begin{figure*}[t] 
  \centering
  \includegraphics[width=\linewidth]{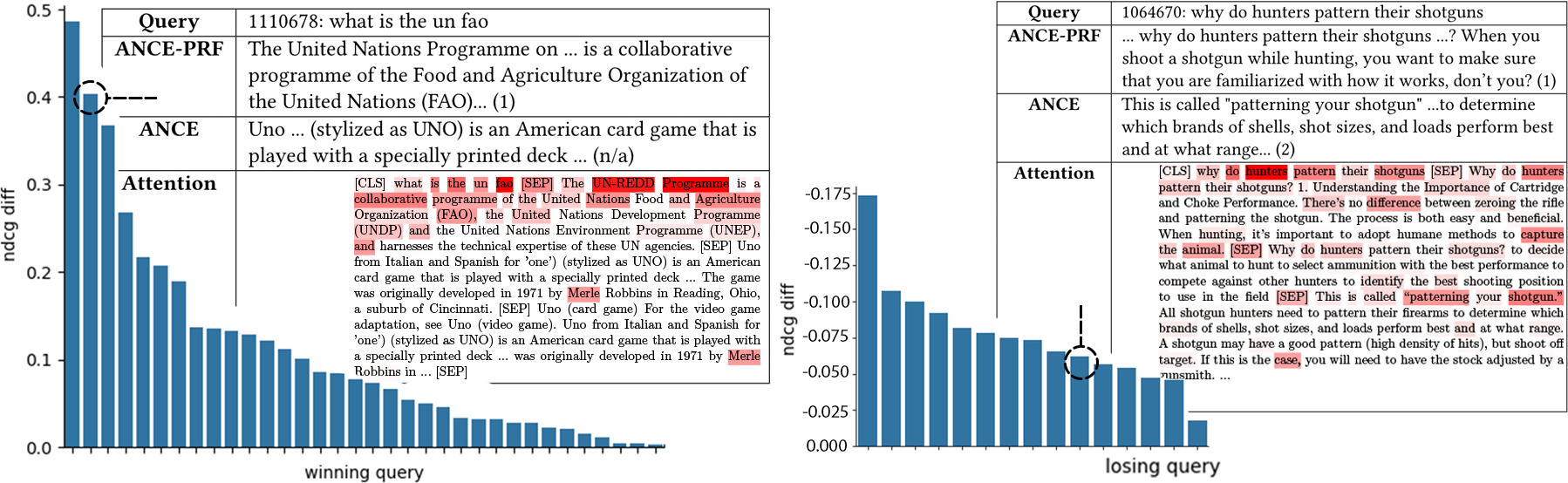}
  \caption{The histograms show the per-query NDCG@10 differences between ANCE-PRF and ANCE retrieval on TREC DL 2020's 54 queries. ANCE-PRF wins on 34 queries, loses on 15, and ties on 5. The tables show the example queries and the two models' first different retrieved passages. Terms receiving higher attention weights are highlighted in darker red.}
  \label{fig:case_study} 
\end{figure*}

\subsection{Ablation on PRF Depths} \label{sec:fb_info}
To understand the number of feedback documents ($k$) needed for effective learning, we trained models using different $k$ and report the results in Table~\ref{tab:ablation}. We trained $k=0$ as a controlled experiment, which is equivalent to training ANCE for an extra 450K steps with fixed negatives. 

Overall, we observe that models with $k>0$ are consistently better than ANCE ($k=\ast$) and $k=0$, showing that ANCE-PRF effectively utilizes the given feedback relevance information. The Avg\_Rel indicates that PRF documents at $k > 1$ contain noisy relevance information, which is a known challenge for traditional PRF approaches~\cite{collins2009reducing}. Nevertheless, ANCE-PRF yields stable improvements over ANCE for $k=1$ to 5, demonstrating the model's robustness against noisy feedback from deeper $k$. 





\subsection{Analyses of Embedding Space \& Attention} \label{sec:dot_prod}
In this group of experiments, we analyze the learned embeddings and attention in ANCE-PRF. 

\textbf{Embedding Space.} Fig.~\ref{fig:dp_attn}(a) shows the distance during training between the ANCE-PRF query embedding and the embeddings of the original ANCE query, the relevant documents, and the irrelevant documents. We use MARCO dev in this study, in which about one out of the three PRF documents is relevant.
In the embedding space, ANCE-PRF queries are closest to the original query and then the relevant documents, while further away from the irrelevant documents.
ANCE-PRF's query embeddings effectively encode both the query and the feedback relevance information. 

\textbf{Learned Attention.} We also analyze the learned attention on the relevant and the irrelevant PRF documents during training. We use TREC DL 2020 for this study as its dense relevance labels provide more stable observations. 
We calculate the average attention from the \texttt{[CLS]} token to each group (``relevant", ``irrelevant", and ``all") of PRF document (Eq. (\ref{eq:cls_attn}) \& (\ref{eq:goal})), and  plot them in Fig.~\ref{fig:dp_attn}(b)-\ref{fig:dp_attn}(d). 



As training proceeds, ANCE-PRF pays more and more attention to the relevant PRF documents than the irrelevant ones, showing the effectiveness of its learning. 
Note that the original query always attracts the most attention from the PRF encoder, which is intuitive, as the majority of the search intent is to be determined by the query. The PRF information is to refine the query representation with extra information but not to invalidate it. 



\subsection{Case Study}
Fig.~\ref{fig:case_study} plots the per query win/loss of ANCE-PRF versus ANCE on TREC DL 2020 and shows one example each.

ANCE-PRF wins on more queries and with larger margins. We also notice the PRF query encoder focuses more on terms that are complementary to the query. In the winning example, ANCE-PRF picks up terms explaining what "un fao" is and does not mistake "un" as "uno".
On the other hand, ANCE-PRF may be misled by information appearing in multiple feedback documents. This is a known challenge for PRF because the correctness of information from multiple feedback documents is its core assumption~\cite{rm12}. 
In the losing example, ``pattern their shotguns" occurs in multiple PRF documents, attracting too much attention to allow ANCE-PRF to make a better choice. 


\section{Conclusion}
Existing dense retrievers learn query representations from short and ambiguous user queries, thus a query representation may not precisely reflect the underlying information need. 
ANCE-PRF addresses this problem with 
a new query encoder that learns better query representations from the original query and the top-ranked documents from a state-of-the-art dense retriever, ANCE. 

Our experiments demonstrate that ANCE-PRF's effectiveness in refining query understanding and its robustness against noise from imperfect feedback. Our studies reveal that ANCE-PRF learns to distinguish between relevant and irrelevant documents. We show that ANCE-PRF successfully learns to identify relevance information with its attention mechanism. Its query encoder pays more attention to the relevant portion of the PRF documents, especially the PRF terms that complement the query terms in expressing the information need. 

ANCE-PRF provides a straightforward way to leverage the PRF information in dense retrieval and can be used as a plug-in in embedding-based retrieval systems. 
We observe that simply leveraging the classic PRF information in the new neural-based retrieval regime leads to significant accuracy improvements, suggesting that more future research can be done in this direction.

\balance
\bibliographystyle{ACM-Reference-Format}
\bibliography{ref}

\end{document}